\documentclass[aps,11pt,epsf]{article}
\usepackage{epsfig}
\title{Microscopic Simulation of Reaction-Diffusion Processes and Applications
to Population Biology and Product Marketing}
\author {Eldad Bettelheim  and Benny Lehmann}
\date{October 1998}
\begin{document}
\maketitle
\roman{figure}
\setcounter{figure}{0}
\newpage
\newpage
\section{Introduction}
Reaction-diffusion processes are the subject of much research 
\cite{fitzhugh} 
\cite{nagumo}
\cite{pearson_9}
\cite{pearson_10}
\cite{gray_8}
\cite{turing_7},
 a reaction-diffusion process occurs as reactants in a solution diffuse in the
liquid and react amongst themselves. A common approach to reaction-diffusion
processes is to consider the density fields of the different reactants
participating in the reactions. This approach stands in contrast to the more
naive approach of tracking the locations of the different reactants, or
computing the wave functions of the different reactants.
Whatever approach is taken the interest in a reaction-diffusion system is
usually in its spatio-temporal evolution. The density field approach is especially adept
for this purpose, since the actual location of specific reactants is, usually
of no interest. In the density fields approach the spatio-temporal evolution is modeled
through partial differential equations (PDE's).

Another approach to reaction-diffusion processes that we have suggested is the
microscopic approach. In this approach we consider the number of reactants at
discrete lattice points, where the lattice models space. The main difference
from the density field approach is that rather than using continuous
densities in a continuous space as in the density field approach, we
use discrete densities in discrete space.

The microscopic simulation approach is closer to the real simulated system when there are only trace
densities of the different reactants. This is because it is in this situation
that the discrete nature of the reactants comes into play. Consequently the PDE
approach describes the system with less accuracy than when there are many reactants.

Reaction-Diffusion processes are not restricted to describing chemical
systems. Indeed reaction-diffusion processes have even been used extensively in
population biology \cite{maynard}. We have also used a reaction-diffusion
model in a marketing context. We have seen that discretization was crucial in
the behavior of the modeled market. Thus showing that microscopic
simulation could be of use to researchers who need to model real-life systems.

\section{The Density Field Approach}

\subsection{Analytical approach}

In this section we shall describe the density field approach \cite{maynard}.
 As
mentioned in the introduction, The density field approach considers the
evolution of the the density fields of the reactants participating in the
reaction-diffusion system. Say that reactants numbered 1 to n are participating in the reactions
(possibly as reactants or as products). Then we could label the density
fields by \(\rho_{i}(\vec{x},t)\). The density is defined as:
\begin{equation}
\label{eq:density_definition}
\rho_{i}(\vec{x},t)=\frac{1}{n_{0}}\lim_{V(A) \to 0}\frac{N(i,A,\vec{x},t)}{V(A)}
\end{equation}

A is a box, \(N(1,A,\vec{x},t)\) is the number of molecules of type \(i\) that are in
the box A located about \(\vec{x}\). \(V(A)\) is the volume of box \(A\). And
\(n_{0}\) is a constant that serves the same purpose as Avogadro's
constant. It should be noted that, if a smooth density function is wanted, 
the limit should not be taken to zero
literally, but rather should be taken down to a scale much larger than one that
shows the discretization of the reactants, and much smaller than the scale of
the macroscopic spatial patterns. 

Let us assume \(l\) possible reactions, where reaction i is of the form:
\begin{equation}
\label{eq:typical_reaction}
S_{j_{1,i}}+...+S_{j_{m_{i},i}}\longrightarrow{}S_{k_{1,i}}+...+S_{k_{n_{i},i}}.
\end{equation}
Where \(S_{r}\) denotes the \(r^{th}\) reactant. As an example of such a
reaction let us look at:
\begin{equation}
\label{eq:example_reaction}
S_{1}+S_{1}+S_{2}\longrightarrow{}S_{1}+S_{1}+S_{1}.
\end{equation}

this reaction is of the form (\ref{eq:typical_reaction}). An interpretation of this reaction is that
two reactants of species 1 can cause a reactant of species 2 to turn into a
reactant of species 1. 
Let us consider a system which has two chemicals, which we shall denote, as
usual, by \(S_{1}\) and \(S_{2}\). These chemicals can diffuse with diffusion
coefficients of \(D_{1}\) and \(D_{2}\) respectively. These chemicals can also
react according to the reaction scheme (\ref{eq:example_reaction}). 

The time evolution of the fields, \(\rho_{1}(\vec{x},t)\)
and \(\rho_{2}(\vec{x},t)\), is given by:
\begin{equation}
\label{eq:time_evol_ex1}
\frac{\partial{}\rho_{1}}{\partial{}t}=D_{1}\nabla^{2}\rho_{1}+k\cdot\rho_{1}^{2}\rho_{2}
\end{equation}

\begin{equation}
\label{eq:time_evol_ex2}
\frac{\partial{}\rho_{2}}{\partial{}t}=D_{2}\nabla^{2}\rho_{2}-k\cdot\rho_{1}^{2}\rho_{2}
\end{equation}

Equation \ref{eq:time_evol_ex1} has two terms on the right hand side (RHS), let us turn our
attention first to the second term. The term contains the expression
\(\rho_{1}^{2}\rho_{2}\). This is proportional to the probability that two
reactants of species 1 and one reactant of species 2 meet in a small region in
space (the volume of that region is a given). The coefficient \(k\) is the
probability that, once the reactants met in the small region in space, they
will react with one another. The Diffusion term is the
familiar term, which originates from the ``random walk'' of the reactants.

In general, when we have \(l\) reactions, each of the form (\ref{eq:typical_reaction}),
\begin{equation}
\label{eq:l_reactions}
S_{j_{1,i}}+...+S_{j_{m_{i},i}}\longrightarrow{}S_{k_{1,i}}+...+S_{k_{n_{i},i}}.
\end{equation}

If we denote by \(\mathbf{N}_{p,r}\)  the number of reactants of species \(\mathbf{r}\)  created or annihilated by reaction \(\mathbf{p}\) then we have the following rate equations:
\begin{eqnarray}
\label{eq:general_rate_PDE}
\frac{\partial\rho_{r}}{\partial{}t}=
\sum_{p=1}^{l}k_{p}N_{p,r}
\prod_{s=1}^{s=m_{p}}\rho_{j_{s,p}}+
D_{r}\nabla^{2}\rho_{r}
\end{eqnarray}

\subsection{Simulation by finite difference}

The former section introduced the formalism of the density field approach
which is essentially analytic, but there is no data structure on a computer
that can hold an arbitrary continuous field. So space is discretized in the computer
simulation. The other problem which arises is the need to integrate the
differential equations over time. This is done again by discretization but the
solution now is to discretize time. The most naive way to integrate using the
differential equation is by Euler integration. This method's main drawback is
the computation time that it requires to get accurate results. But
fundamentally it is no different than other more sophisticated methods such as
the runga-cutta method. We shall outline this finite difference approach using
Euler integration below.

In the finite difference approach using Euler integration one replaces the
fields \(\rho_{r}(\vec{x},t)\) with
\(\vec{x}\in{}R^{d}\) and \(t\in{}R\) by
\(\rho_{r}^{*}(\vec{x},t)\) where
\(\vec{x}\in{}\omega^{d}\) \(t\in{}\omega\) (\(\omega\)
being the natural numbers). Let us suppose that the following equations hold
for the fields \(\rho_{r}\):
\begin{equation}
\label{eq:rate_equals_f}
\frac{\partial\rho_{r}}{\partial{}t}=f(\rho_{1},...,\rho_{n})+D\nabla^{2}\rho_{r},
\end{equation}

In order to make this transition from \(\rho\) to
\(\rho^{*}\), space is conceptually divided to a discrete d-dimensional lattice of spacing
\(\Delta{}x\) and time is divided to a discrete 1-dimensional lattice (actually
a series) of spacing \(\Delta{}t\). Now the aim is to
make the following equality be a good approximation: 
\begin{equation}
\label{eq:finite_differ_approx}
\rho_{r}^{*}\left(\left(n_{1},n_{2},...,n_{d}\right),m\right) 
\approx{}
\rho_{r}\left(\Delta{}x\cdot\left(n_{1},n_{2},...,n_{d}\right),m\cdot{}\Delta{}t\right)
\end{equation}

The way to make this approximation good is to let \(\Delta{}x\) and
\(\Delta{}t\) be small and to let \(\rho^{*}\) follow the dynamics:
\begin{eqnarray}
\label{eq:dynamics_finitedif_f}
\rho_{r}^{*}(\vec{x},t+1)-\rho_{r}^{*}(\vec{x},t)=\Delta{}t\cdot{}f(\rho_{1}^{*},...,\rho_{n}^{*})+\frac{D_{r}\Delta{}t}{\Delta{}x^{2}}
\nonumber\\
(\rho_{r}^{*}({\vec{x}}+(1,0,...0),t)+\rho_{r}^{*}({\vec{x}}+(-1,0,...0),t)+\nonumber\\
\rho_{r}^{*}({\vec{x}}+(0,1,...0),t)+\rho_{r}^{*}({\vec{x}}+(0,-1,...0),t)+ ...\nonumber\\
\rho_{r}^{*}({\vec{x}}+(0,0,...1),t)+\rho_{r}^{*}({\vec{x}}+(0,0,...-1),t)
-2d\rho_{r}^{*}({\vec{x}},t))
\end{eqnarray}
The first term on the right hand side of the equation is the first order
approximation of the difference between \(\rho_{r}(\vec{x},t)\)
and  \(\rho_{r}(\vec{x},t+\Delta{}t)\), after a time
interval of \(\Delta{}t\) has passed assuming the dynamics
(\ref{eq:rate_equals_f}). 
The second term accounts for diffusion
and includes the discretization of the \(\nabla^{2}\) operator. This term includes
positive and negative terms. The positive terms are contributions to the density
at site \(\vec{x}\)  from densities at neighboring
sites. Neighboring sites are those sites which have the same coordinates as
\(\vec{x}\) but for a single coordinate which must be only
one lattice point away. This contribution is due the fact that diffusion
causes chemicals to move from one location to another in space. The negative
term accounts for the chemicals leaving site  \(\vec{x}\).

Now, if we replace \(f\) with the terms from the density field approach to
reaction-diffusion, we get:
\begin{eqnarray}
\label{eq:full_finite_difference}
\rho_{r}^{*}(\vec{x},t+1)-\rho_{r}^{*}(\vec{x},t)=\nonumber\\
\Delta{}t\sum_{p=1}^{l}k_{p}N_{p,r}
\prod_{s=1}^{s=m_{p}}\rho_{j_{s,p}}^{*}(\vec{x},t)
\nonumber\\
+\frac{D_{r}\Delta{}t}{\Delta{}x^{2}}
(\rho_{r}^{*}({\vec{x}}+(1,0,...0),t)+\rho_{r}^{*}({\vec{x}}+(-1,0,...0),t)+\nonumber\\
\rho_{r}^{*}({\vec{x}}+(0,1,...0),t)+\rho_{r}^{*}({\vec{x}}+(0,-1,...0),t)+ ...\nonumber\\
\rho_{r}^{*}({\vec{x}}+(0,0,...1),t)+\rho_{r}^{*}({\vec{x}}+(0,0,...-1),t)
-2d\rho_{r}^{*}({\vec{x}},t))
\end{eqnarray}

\section{Microscopic Simulation of Reaction-Processes}
\subsection{Fundamentals of the Approach}
In the previous section we have seen the prevailing approach for dealing with
reaction-diffusion processes. Another approach that can be used is the
microscopic simulation approach. This approach takes discretization one step
further in the sense that the fields are discretized themselves, but takes a
welcomed step backwards in the sense that time is not discretized. This
approach  is useful because it is closer to reality. Chemicals
are discrete entities (at least in the classical approach which is a good
approximation in solutions). 

Again we have a field \(\rho_{r}^{**}(\vec{x},t)\) where
\(\vec{x}\in{}\omega^{d}\) \(t\in{}\omega\),
but this time \(\rho_{r}^{**}(\vec{x},t)\in\omega\). We
have said that time is not considered discrete in the microscopic simulation
model, but nevertheless we have \(t\in{}\omega\). This is not a contradiction
it is simply an expression of the fact that reactions occur at discrete time
points. Let us denote simulation discrete time with \(t^{*}\) and real time
with \(t\). Say the reactions occur, in  the real system, at times \(t_{n}\).
Then when the simulation is at time \(t^{*}=n\) it is supposed to approximate
the real system at time \(t_{n}\). Simulating the real system between the times
\(t_{n}\) is of no use since nothing happens, except for diffusion which we
should also treat as a reaction. But the problem is that diffusion, in the real system, is not a
process that occurs at time points, rather it is a continuous process. On the
other hand we have modeled space by discrete sites. Diffusion is modeled by
chemicals hopping from one site to another. This process is discrete since chemicals
hop at discrete time points. So amongst the times \(t_{n}\) there are times
at which the reaction which takes place is diffusion, that is to say hopping of chemicals
to neighboring sites. We should stress that the time interval between
\(t_{n}\) and \(t_{n+1}\) is not a constant. So the real time is not approximated by
\(t^{*}\cdot\Delta{}t\) for some \(\Delta{}t\).

The time interval between \(t_{n}\) and \(t_{n+1}\) is large
when the time interval between successive reactions, in the real system, is
large. Roughly speaking this happens when there are not many reactants, or many
inert reactants.  This is also when 
the microscopic simulation is at its best (in terms of the simulation's speed),
 since the simulation's
single step covers a lot of time. We shall give a quantative estimate for this
time interval later.

Let us now turn to the relation of \(\rho^{**}\) to the real system that it is
supposed to approximate. We assume space of dimensions d. Let us denote by  
\(N_{r}(\vec{x},l,t)\) the
number of reactants, in the real system, of species \(r\), located in a box of
length \(l\) around \(\vec{x}\) at time \(t\). The
approximation relation is given below:
\begin{equation}
\label{eq:mic_react_approx}
\rho^{**}_{r}(\vec{x},t^{*})
\approx{}N_{r}(\Delta{}x\cdot\vec{x},\Delta{}x,t_{t^{*}})
\approx{}\rho^{*}_{r}(\vec{x},t_{t^{*}})\cdot\Delta{}x^{d}n_{0}
\end{equation}

Let us imagine a grid of spacing \(\Delta{}x\) dissecting the real system, so
that space is divided into little boxes. This division is not physical, but
mental. Now each such box is simulated as a site on the simulation lattice. The
number of reactants at each lattice point should approximate the number of
reactants in a the little boxes imagined in the real system. This is the nature
of the first approximate equality in equation (\ref{eq:mic_react_approx}) 
. The second approximate equality in equation
(\ref{eq:mic_react_approx}) is due to
the approximation of the finite difference approach to the real system.

\subsection{The Monte-Carlo Method}
Now we shall introduce the dynamics of
\(\rho^{**}\). Since the microscopic simulation is a Monte-Carlo simulation,
\(\rho^{**}\)'s dynamics follow the following rules:
\begin{enumerate}
\item{} Choose \(\rho^{**}(\vec{x},t^{*}+1)\) from a
probability space, \(\Omega_{0}(t^{*})\).
\item{} Calculate the new probability space \(\Omega_{0}(t^{*}+1)\).
\end{enumerate}

\(\Omega_{0}(t)\) associates a probability  for  every possible
\(\rho^{**}(\vec{x},t+1)\),  but actually we are intent
on performing one reaction at a time. So \(\Omega_{0}(t)\) will give non-zero
probabilities only for those \(\rho^{**}(\vec{x},t+1)\)
that differ from \(\rho^{**}(\vec{x},t)\) by a single
reaction (or diffusive hopping). So we can look at \(\Omega_{0}(t)\) as associating a probability  for
every possible reaction. So let us construct a probability space \(\Omega(\Re,t)\)
which associates a probability for each possible reaction.

Let us speak of a reaction, \(\Re\). This reaction can potentially take place
anywhere in real space. In particular the reaction can fall within any one of the
little boxes that we have discussed in the previous sub-section.
We shall denote reaction \(\Re\) taking place at a little box corresponding to 
site \(\vec{x}\) by \(\Re_{\vec{x}}\).
Our task
now is to find out, given some initial conditions, what is the probability that
the reaction that will take place next is \(\Re_{\vec{x}}\).

Let us expand a bit on the stochastic process that the real system
undergoes. The stochastic process is comprised of events (reaction and
diffusion) occurring stochastically at
discrete time points. The events we are considering are the real
reactions and the movement of reactants from one little box to an adjacent
one. Let \(dt\) be a small time interval, then for \(\Re_{\vec{x}}\) there is a chance
\(P(\Re_{\vec{x}},t^{*})dt\) that this reaction will occur in the time interval \(dt\). The
probability density, \(P(\Re_{\vec{x}},t^{*})\),
 is called the reaction rate for reaction \(\Re_{\vec{x}}\), and it is exactly
what we used in order to formulate the PDE for the real system, as we shall
soon see. First let us notice that this probability density has reciprocal time
as units, which is consistent with the name 'rate'.

The connection of \(P(\Re_{\vec{x}},t^{*})dt\) to the PDE's will be useful in calculating
\(P(\Re_{\vec{x}},t^{*})\). 
So let us explore this connection. Reaction \(\Re\) is of the
form (1), that is to say: 
\begin{equation}
\label{eq:reaction_re}
S_{j_{1,\Re}}+...+S_{j_{m_{\Re},\Re}}\longrightarrow{}S_{k_{1,\Re}}+...+S_{k_{n_{\Re},\Re}}.
\end{equation}

A crucial assumption for the
validity of the PDE's is that there is some time scale, \(dt\), during which
\(P(\Re_{\vec{x}},t^{*})\) does not change much and still for the same time scale, \(dt\), many
reactions occur. Under these assumptions it can be shown that the number of
reaction and diffusion events that occur at the time interval \(dt\) in the box
around \(\vec{x}\) is, simply, \(P(\Re_{\vec{x}},t^{*})dt\). Let us
look at species \(r\) ( recall the notation used in equation
(\ref{eq:reaction_re}) ). Now let\footnote{\#S denotes the number of elements
in S.}
\begin{equation}
\label{eq:s_imbalance}
N_{\Re,r}=\#\{n|n\leq{}n_{\Re},k_{n,\Re}=r\}-\#\{n|n\leq{}m_{\Re},j_{n,\Re}=r\}.
\end{equation}

This \(N_{\Re,r}\) gives the number of reactants of species \(r\) created in the reaction
\(\Re\) ( a negative number indicates that the species is annihilated in the
reaction). So the number of reactants of
species \(r\) formed, at the box around \(\vec{x}\), in the time interval \(dt\) is:
\begin{equation}
\label{eq:formed_interval}
P(\Re_{\vec{x}},t^{*})\cdot{}dt\cdot{}N_{\Re,r}
\end{equation}

The number of reactants of species \(r\) at the box which
corresponds to site \(\vec{x}\) at time \(t_{n}\)  is approximated by
\(\rho^{**}_{r}(\vec{x},n)\). So the rate at
which \(\rho^{**}_{r}(\vec{x},n)\) changes due to reaction \(\Re_{\vec{x}}\)
is given by:
\begin{equation}
\label{eq:rate_of_reaction}
\frac{\partial\rho^{**}_{r}}{\partial{}t}_{\Re_{\vec{x}}}=P(\Re_{\vec{x}},t^{*})\cdot{}N_{\Re,r}.
\end{equation}

Where the index \(\Re_{\vec{x}}\) denotes that the reference is to the rate of change due
to reaction \(\Re_{\vec{x}}\) alone. We have already seen the rate of change in
the finite difference case (equation (\ref{eq:full_finite_difference})). There we expressed
\(\frac{\partial\rho^{*}}{\partial{}t}\) as a sum of terms, each expressing a
rate due to different reactions. Another term was due to diffusion. Realizing
that the terms appearing in the finite difference case express the same thing
as the rate expressed at (\ref{eq:rate_of_reaction}) modulo the approximation
relation (\ref{eq:mic_react_approx}) we can find an expression for \(P(\Re_{\vec{x}},t)\), this
is given by:
\begin{equation}
\label{eq:expression_for_P}
P(\Re_{\vec{x}},t^{*})=\Delta{}x^{d}{}n_{0}\cdot{}k_{\Re}\prod_{s=1}^{s=m_{\Re}}\left(\rho^{**}_{j_{s,\Re}}\frac{1}{\Delta{}x^{d}{}n_{0}}\right)
\end{equation}

This assumes that the reaction in question is a real reaction, that is not
diffusion. For the case of diffusion of species \(l\) we have the following equation:
\begin{equation}
\label{eq:Pexpression_diffusion}
P(\Re_{\vec{x}},t^{*})=2d\cdot{}D_{l}\cdot\frac{\rho^{**}_{l}}{\Delta{}x^{d}}
\end{equation}

This equation is derived considering the last term in equation
(\ref{eq:full_finite_difference}), 
\(-\frac{2dD_{r}\Delta{}t}{\Delta{}x^{2}}\rho_{r}^{*}(\vec{x},t)\), which is due to
reactants hopping from site \(\vec{x}\) to neighboring sites. And then
considering that \(P(\Re_{\vec{x}},t^{*})\), in equation
(\ref{eq:Pexpression_diffusion}),  is the probability density for
hopping to neighboring sites. After using the approximation relation
(\ref{eq:mic_react_approx}) we get (\ref{eq:Pexpression_diffusion}).

Now that we have an expression for the rates \(P(\Re_{\vec{x}},t^{*})\) of the
different reactions, we still face the task of finding the probability
associated with each possible reaction (including diffusion) by
\(\Omega\). Again let us point out that the probability that \(\Omega\)
associates with reaction \(\Re_{\vec{x}}\) at time \(t^{*}\) is the probability that this
reaction will come next in the sequence of reactions. Now between the times
\(t_{t^{*}}\) and \(t_{t^{*}+1}\) nothing happens in the reaction chamber,
apart from reactants moving inside the little boxes we have imagined. In this
time interval the reactants don't cross the boundaries of the boxes. If the rate of
\(\Re_{\vec{x},1}\) is twice that of 
\(\Re_{\vec{x},2}\) then we should
expect that reaction \(\Re_{\vec{x},1}\) has twice the chance to be the next reaction
that occurs than \(\Re_{\vec{x},2}\). Let us expand a bit on the nature of the
assumption we made in the previous statement. We assume that the probability
distribution associated with \(\Omega\) is dependent only on the situation of
the current configuration of the reaction chamber and is independent of the
time that passed since the last reaction-diffusion event. So it is
actually this assumption (that the stochastic process has no memory) that allows us to compute the probability of different
reaction and diffusion events associated by \(\Omega\) as a function of the rates. 
The preceding  argument leads to the following relation, given two
reaction-diffusion events \(\Re_{\vec{x},1}\) and \(\Re_{\vec{x},2}\):
\begin{equation}
\label{eq:one_P_all_P}
\frac{\Omega(\Re_{\vec{x},1},t^{*})}{\Omega(\Re_{\vec{x},2},t^{*})}=\frac{P(\Re_{\vec{x},1},t^{*})}{P(\Re_{\vec{x},2},t^{*})}
\end{equation}

Taking into account the normalization of probability distribution to one, we now
can compute the probability associated by \(\Omega\) to a reaction-diffusion
event \(\Re_{\vec{x}},t^{*}\).
The solution is:
\begin{equation}
\label{eq:finally_omega}
\Omega(\Re_{\vec{x},j},t^{*})=\frac{P(\Re_{\vec{x},j},t^{*})}{\sum_{\Re_{\vec{x},i}}{P(\Re_{\vec{x},i},t^{*})}}.
\end{equation}

\section{Anderson Localization in a Reaction-Diffusion System}
\subsection{The Reaction-Diffusion System}
Anderson localization is a phenomenon associated with electron-transport
behavior in disordered materials. The disorder in the material induces a phase
transition of the electron eigen-functions from an unlocalized state in which Ohm's
law is valid into a localized state in which the material behaves as an insulator.
The wave functions of an electron under the influence of a
periodic potential is periodic. A question arises, in the context of
 disordered materials,
 of how this periodic
wave function is influenced by disturbances to the periodicity of the
potential. One might think that the eigenfunction of a slightly perturbed
potential (that is to say perturbed from an originally periodic potential),
would be eigen-fucntions slightly perturbed from a periodic function. This
intuition proves misleading in the metal-insulator transition case. It is seen that some of
the eigen-functions exhibit a marked departure from periodic functions, even for
small disturbance of the potential. These functions are seen to be
localized. Meaning that there are small ``islands'' in which the wave function
has high modulus and there are large spaces between these islands where the wave-function has low
modulus. Indeed the wave functions that depart from periodicity have the
approximate form :
\begin{equation}
\label{eq:localized_state}
e^{-\frac{\left(x-x_{0}\right)}{\xi}}
\end{equation}

\(\xi\) is the localization length of the eigenfunction. This
approximation is good for the tails of the eigenfunction.

This Anderson localization effect is also observed in reaction diffusion
system. Let us take the example given by Shnerb and Nelson. They presented a 
reaction-diffusion system described the schematics:
\begin{eqnarray}
\label{eq:anderson_schematics}
A+F\longrightarrow{}A+A+F
\nonumber\\
A+A\longrightarrow{}\emptyset
\end{eqnarray}

\(\mathbf{A}\) is the only reactant to undergo diffusion.\\
Where \(\emptyset\) signifies that the reaction has no products (alternatively
it can be interpreted that the products of the reaction are
inert). These reactions together with diffusion can be seen as the schematics
of the population biology of bacteria. The first reaction accounts for the
reproduction of the bacteria. The rate at which the bacteria reproduce is
controlled also by the concentration of F. F can represent, for example, the intensity of
light that falls on the bacteria \cite{shnerb_1}. But can also represent
any factor that controls the rate of bacteria reproduction, with the provision
that this factor is not
variable in time, and in particular non-exhaustible. The second reaction  accounts for the
dying of bacteria due to overcrowding. The reaction dynamics has the species
\(\mathbf{A}\)
 facilitating the production of more \(\mathbf{A}\)'s. This property
\(\mathbf{A}\) is called autocatalysis. Autocatalysis is an important concept in pattern
formation \cite{turing_7}, the origins of life
\cite{dyson}
\cite{segre}
\cite{kauffman_model}
\cite{kauffman}
\cite{oparin}
and economy \cite{sorin_5} (where autoctalysis is the underlying concept of
multiplicative dynamics).

To see the similarity of this problem to an eigenfunction problem in quantum
mechanics, such as Anderson localization, let us write down the PDE's which are
associated with the reaction dynamics above:
\begin{equation}
\label{eq:full_shnerb}
\frac{\partial{}\rho_{A}}{\partial{}t}=D_{A}\nabla^{2}\rho_{A}+\rho_{A}\cdot{}\rho_{F}-\rho_{A}^{2}
\end{equation}

which can be reformulated into:
\begin{equation}
\label{eq:full_near_operator}
\frac{\partial{}\rho_{A}}{\partial{}t}=(D_{A}\nabla^{2}+\rho_{F}-\rho_{A})(\rho_{A})
\end{equation}

Now if we drop the last term, we have:
\begin{equation}
\label{eq:linear_near_operator}
\frac{\partial{}\rho_{A}}{\partial{}t}=(D_{A}\nabla^{2}+\rho_{F})(\rho_{A})
\end{equation}

which is quite analogous to the time dependent Schr\"{o}dinger equation:
\begin{equation}
\label{eq:schrodinger}
i\frac{\partial{}\psi}{\partial{}t}=(\frac{1}{2m}\nabla^{2}+U(x,t))(\psi)
\end{equation}

Where we have \(U\) playing the part of \(\rho_{F}\), \(\frac{1}{2m}\)
playing the part of \(D_{A}\) and \(\psi\) playing the part of \(\rho_{A}\). We
have only to remember that we have dropped the quadratic term and that there is
an additional  coefficient, -\(i\), in the Schr\"{o}dinger equation \footnote{\(i\) here
is \(\sqrt{-1}\)}. Consequently, the reaction-diffusion PDE, without the
quadratic term, can be seen as a Schr\"{o}dinger equation with imaginary time. 

Notice that we did not write an equation for
\(\frac{\partial\rho_{F}}{\partial{t}}\) since \(\rho_{F}\) does not undergo
any dynamics and therefore does not change in time. The phenomena of Anderson localization
determines that if \(\rho_{F}\) is not constant in space, but rather is
stochastic, then we shall have these localized states that we have described
above, both for the quantum mechanics case and for the reaction-diffusion
case. In the quantum mechanics case, a constant or periodic potential,
\(U(x,t)\), entails a periodic field, \(\psi\). In the reaction-diffusion
case, a constant potential, \(\rho_{F}\), entails a constant field,
\(\rho_{A}\). 

A constant, stochastic potential naturally arises when speaking of
disordered materials \cite{efros}. But in reaction-diffusion systems, such a potential is
more of a stretch. We have given the example of light intensity as a stochastic
constant potential. But in many more cases the reaction rate associated with
the reproduction of bacteria, or some chemical, is dynamic.
This leads us to deal with a dynamic potential, \(U(x,t)\). Nelson and
Shnerb have already discussed a time-dependent potential of the form
\(U(x-vt)\) (actually they have looked at the case where the media is moving,
which is equivalent to a moving potential in the opposite direction).

\subsection{Results of microscopic simulations for Anderson localization}

We simulated (using microscopic simulation) the reaction dynamics associated with
Anderson localization . 
The reaction schematics (together with the reaction rates) is given
below\footnote{The numbers in parentheses at the right of the reactions denote
their corresponding rates}:

\begin{enumerate}
\item \(\mathbf{A} + \mathbf{C} \rightarrow  \mathbf{A}  + \mathbf{A} + \mathbf{C}  (\mathbf{15})\)
\item \(\mathbf{A} + \mathbf{A} \rightarrow \emptyset (\mathbf{3})\)
\item \(\mathbf{A} \rightarrow \emptyset (\mathbf{60})\)
\end{enumerate}

\begin{figure}
\frame{
\vspace {0.01in}
\epsfxsize=3.1in
\epsfbox{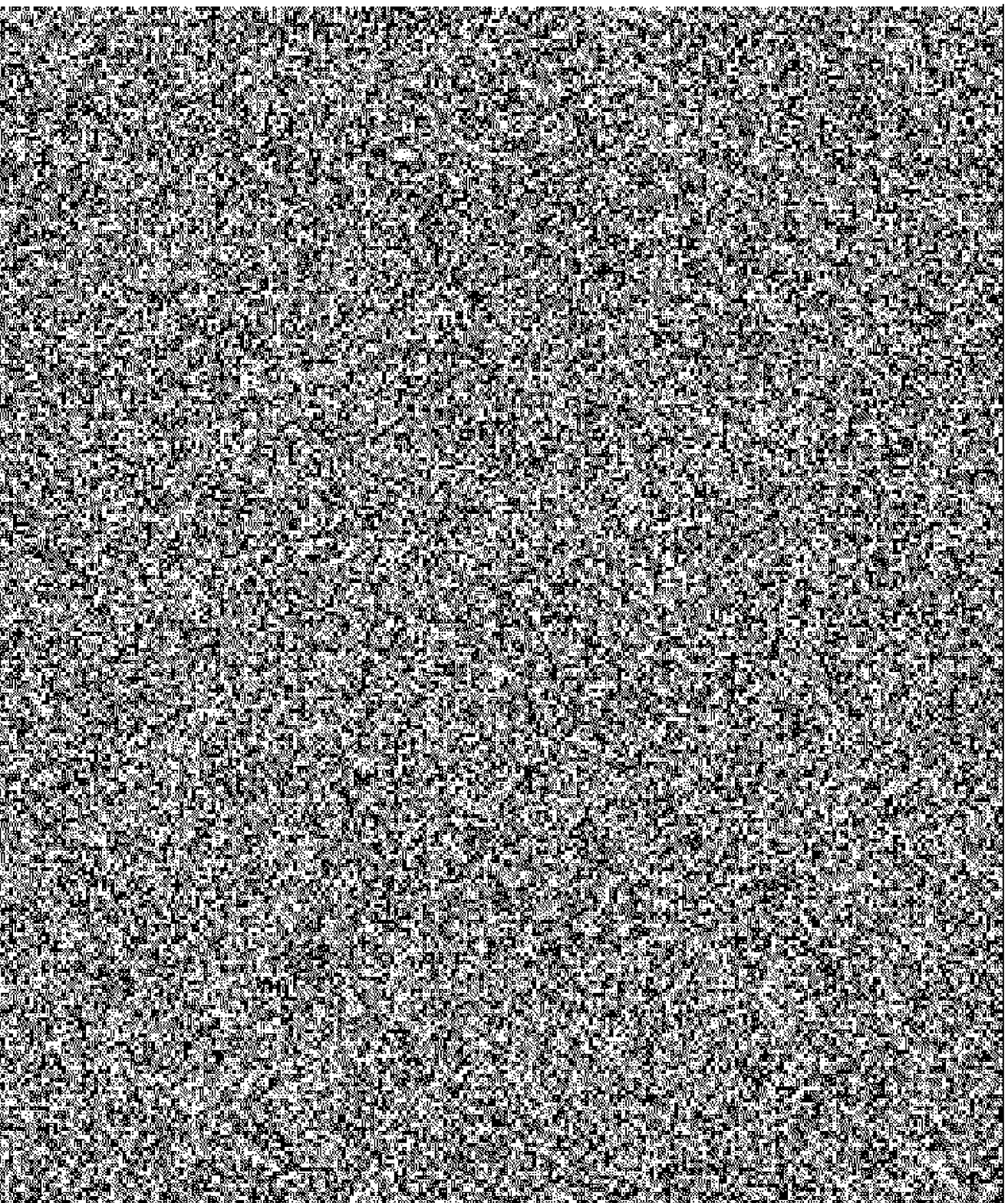}
}
\caption{
Snapshot of molecules of type \(\mathbf{C}\). 
}
\label{fig:Anderson_food}
\end{figure}

\begin{figure}
\frame{
\vspace {0.01in}
\epsfxsize=3.1in
\epsfbox{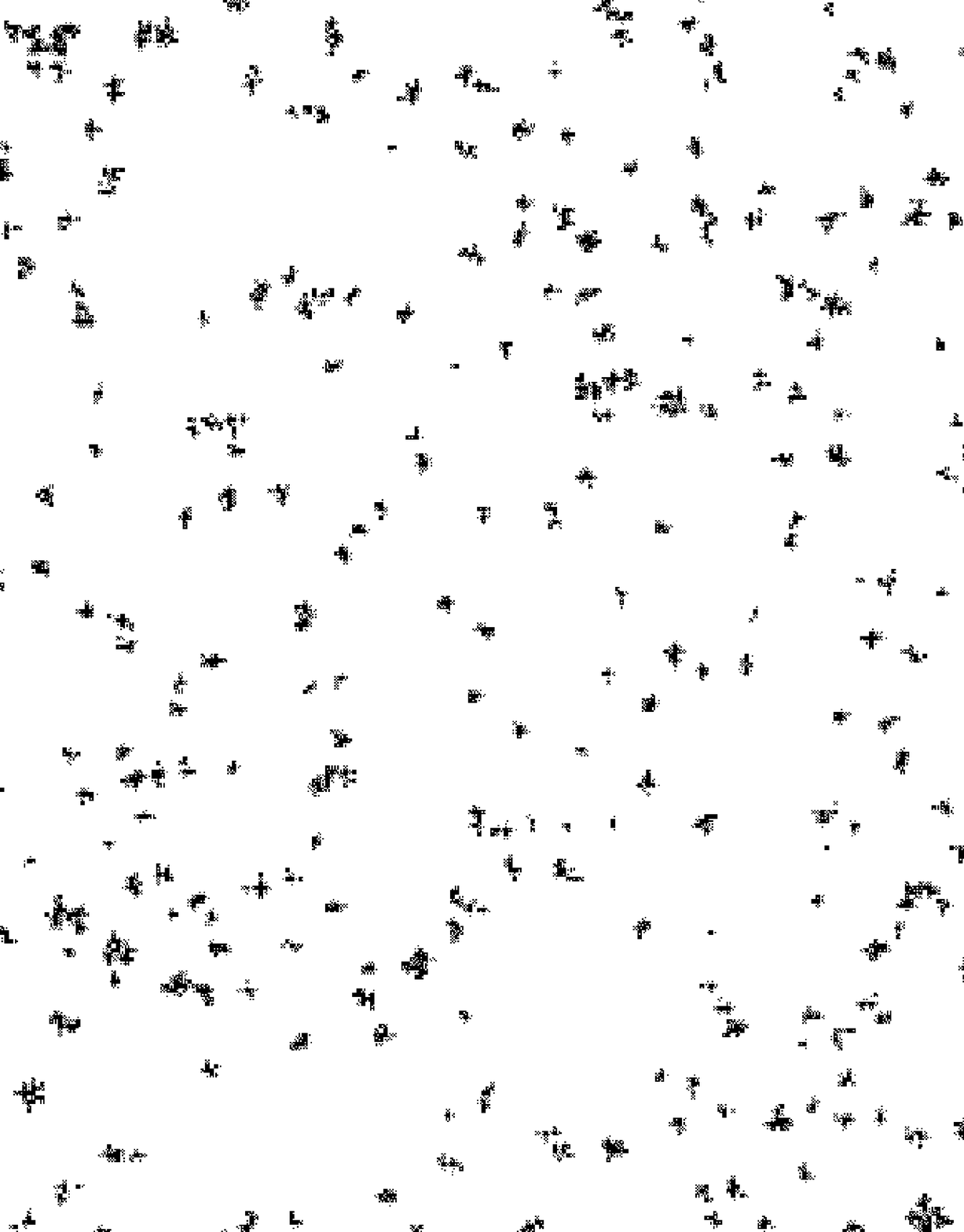}
}
\caption{
Snapshot of molecules of type \(\mathbf{A}\). 
}
\label{fig:Anderson_localization}
\end{figure}

The size of the reaction chamber is 300x300 lattice points. The average number of
molecules of type \(\mathbf{C}\) per site is 5. They are dispersed in the beginning of the simulation
randomly. That is to say that the positions of the 5x300x300 molecules of type 
\(\mathbf{C}\)
are chosen at
random. The spatial distribution that resulted can be seen in figure (\ref{fig:Anderson_food}).
The initial distribution of \(\mathbf{A}\) is similarly dispersed but with and
average of 2 molecules of type \(\mathbf{A}\) per site. Only  \(\mathbf{A}\) molecules diffuse. The simulation's diffusion rate is 3. After awhile a steady state is
reached. Off-course \(\mathbf{C}\) is not dynamic, and it remains as in figure
(\ref{fig:Anderson_food}). On the other hand  \(\mathbf{A}\) is dynamic and the
steady state is depicted in figure (\ref{fig:Anderson_localization})

\subsection{Anderson Localization - in the search of dynamic clustering }
 Shnerb and Nelson\cite{shnerb_1} have explored the consequences of changing the potential
 \(U(x)\)  by a moving potential \(U(x-vt)\), what would happen if instead of
 having the potential drift at a constant speed the potential would undergo
 diffusion itself? This would correspond to replacing (\ref{eq:full_shnerb}) by
 the two coupled equations:
\begin{eqnarray}
\label{eq:potential_diffuse}
\frac{\partial{}\rho_{A}}{\partial{}t}=D_{A}\nabla^{2}\rho_{A}+\rho_{A}\cdot{}\rho_{F}-\rho_{A}^{2}
\nonumber\\
\label{eq:moving_potential}
\frac{\partial{}\rho_{F}}{\partial{}t}=D_{F}\nabla^{2}\rho_{F}
\end{eqnarray}

We have found no treatment in the literature for this kind of moving potential for
a good reason: if \(\rho_{F}\) is treated as non-negative continuous variable
as is usually done in the context of differential equations and since the only
dynamics imposed on \(\rho_{F}\) by (\ref{eq:potential_diffuse}) is diffusion,
\(\rho_{F}\) converges to a steady spatially-uniform state \(\rho_{F}=const\),
giving rise to the trivial solutions of a system with no potential fluctuations
at all. But if the potential is treated as a discrete variable diffusion does
not necessarily induce the uniform distribution of the potential and
localization effects still have a chance to prevail. This situation leads to
clear differences between the different simulation approaches we have
described, in further sections we will show other examples in which
discretization leads to different results in different simulation methods.

Extending the analogy to population biology we could look at \(\mathbf{A}\) as representing
a population of parasites dependent on a host species \(\mathbf{F}\) for a-sexual
reproduction. The population represented by \(\mathbf{F}\) is not affected by
the parasites and performs random diffusion in space. The phenomena
we are most interested in is dynamic clustering or grazing. Dynamic clustering would under
our analogy represent parasite herds moving in space due to changes in the
spatial distribution of the species they need in order to reproduce.
     
Translating our situation into a reaction-diffusion system will give the same
results as (\ref{eq:anderson_schematics}) but in this case both \(\mathbf{A}\)
and \(\mathbf{F}\) undergo diffusion. We have added to
(\ref{eq:anderson_schematics}) the reaction:
\begin{equation}
\label{eq:A_dying}
A\longrightarrow{}\emptyset
\end{equation}
representing the dying of \(\mathbf{A}\) not due to overcrowding.

Since (\ref{eq:anderson_schematics})
describes no creation or elimination of \(\mathbf{F}\) the total number of
\(\mathbf{F}\)s in all the lattice sites is constant throughout the simulation and 
 \(<\mathbf{F}>\) - the average concentration of \(\mathbf{F}\) is
constant as well. Given a constant value for \(D_{A}\) the results of
simulations of (\ref{eq:move_F}) are controlled by \(<\mathbf{F}>\) and \(D_{F}\)
 We have simulated (using microscopic simulation) the following
reaction-diffusion system with different values of \(<\mathbf{F}>\) and
\({D_{F}}\) and keeping \(D_{A}=6\):
\footnote{Numbers in parentheses at the right of reactions are the
corresponding reaction-rates.}
\begin{eqnarray}
\mathbf{A} + \mathbf{F} \rightarrow  \mathbf{A}  + \mathbf{A} + \mathbf{F} (\mathbf{13})
\nonumber\\
\mathbf{A} + \mathbf{A} \rightarrow \emptyset (\mathbf{0.01})
\nonumber\\
\mathbf{A} \rightarrow \emptyset (\mathbf{8})
\nonumber\\
\label{eq:move_F}
\end{eqnarray}

The size of the simulation is 128x128. The system was seeded with 3 reactants
of species \(\mathbf{F}\) per cell and 1 reactant of species \(\mathbf{A}\) per
cell (the reactants were placed at random locations).

\begin{figure}
\frame{
\vspace {0.01in}
\epsfxsize=3.1in
\epsfbox{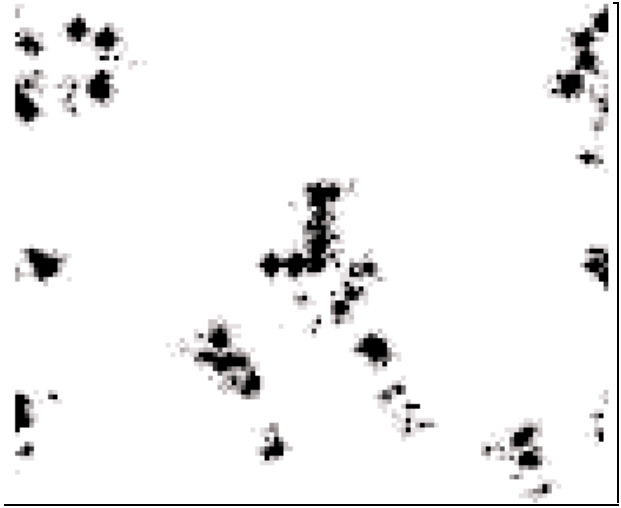}
}
\caption{
Snapshot of \(\mathbf{A}\)'s concentration, simulation parameters are
\(D_{A}=6\),\(D_{F}=4\),\(<\mathbf{F}>\)=0.15) .
}
\label{fig:mesh159a}
\end{figure}

\begin{figure}
\frame{
\vspace {0.01in}
\epsfxsize=3.1in
\epsfbox{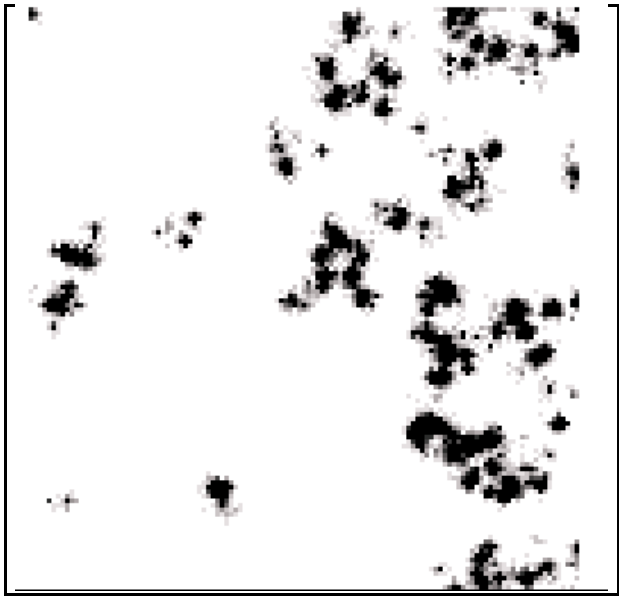}
}
\caption{
Snapshot of the same simulation as in (\ref{fig:mesh159a}) at a later time,
the clusters have moved. The result of this simulation falls into the \(\beta\)
category in our notation.
}
\label{fig:mesh159b}
\end{figure}

We have divided the results of the simulation into 3 categories:
\begin{itemize}
\item\(\alpha\)- The simulation results in \(\mathbf{A}\) filling the whole
simulation space.
\item\(\beta\)- The simulation results in \(\mathbf{A}\) forming dynamic clusters.
\item\(\gamma \)-The simulation results in the total extinction of
\(\mathbf{A}\) from the simulation space.
\end{itemize}

\begin{figure}
\frame{
\vspace {0.01in}
\epsfxsize=3.1in
\epsfbox{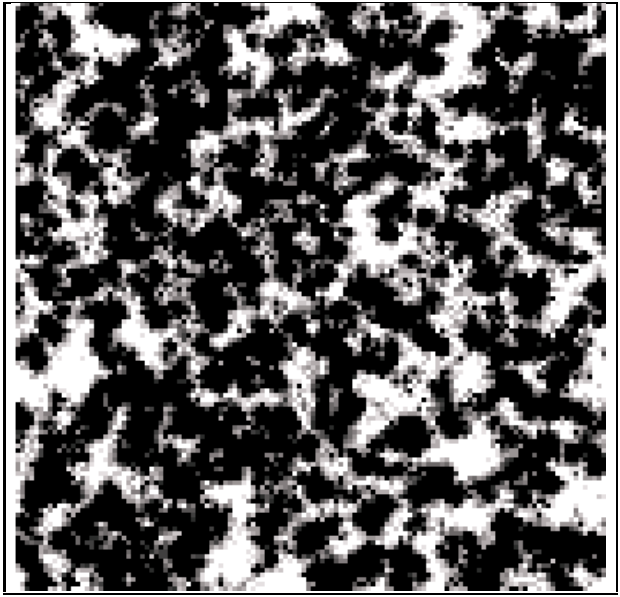}
}
\caption{
\(\mathbf{A}'s\) fill up the simulation space, this corresponds to a \(\gamma\)
situation in
our notation.
}
\label{fig:fillup}
\end{figure}

\begin{figure}
\frame{
\vspace {0.01in}
\epsfxsize=3.1in
\epsfbox{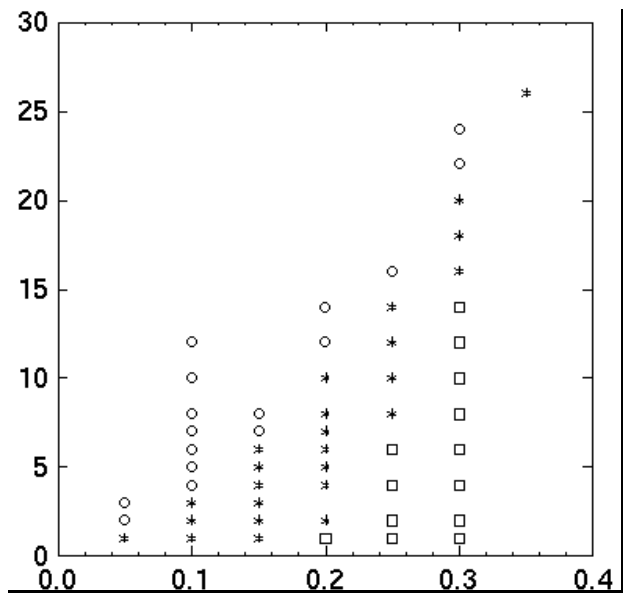}
}
\caption{
Results of simulations of the system(\ref{eq:move_F}) for different values of
\(<\mathbf{F}>\) (x-axis) and \(D_{F}\)(y-axis). Notice that dynamic
clustering occurs also in the realistic range of: \(0.5D_{A}<D_{F}<1.5D_{A}\).
Circles denote simulation resulting in \(\alpha\) situations, Asterisks denote
simulations resulting in  \(\beta\) situations
and squares denote simulations resulting in \(\gamma\) situations.
}
\label{fig:phase}
\end{figure}

\begin{figure}
\frame{
\vspace {0.01in}
\epsfxsize=3.1in
\epsfbox{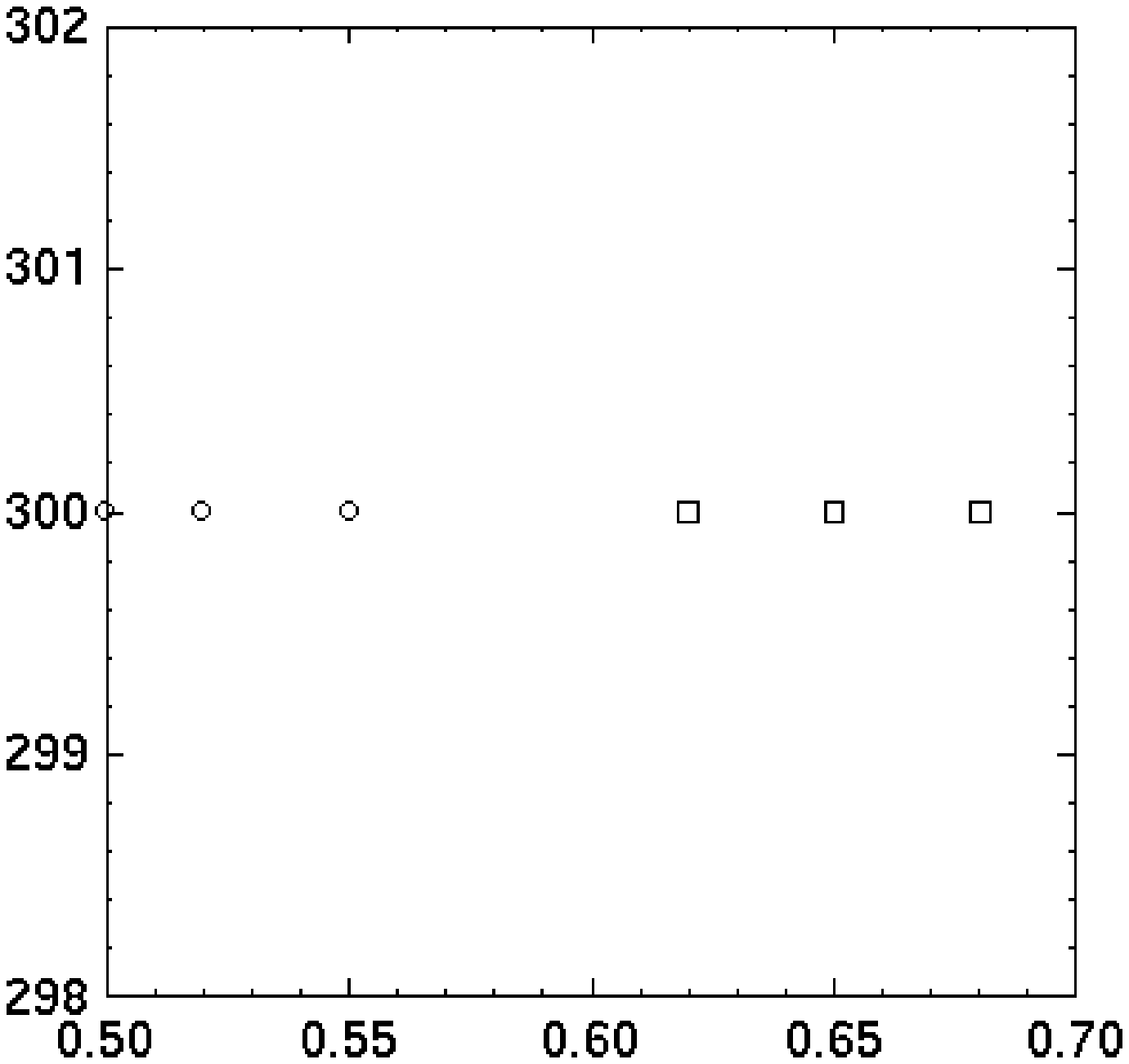}
}
\caption{
Results of simulations of the system(\ref{eq:move_F}) for \(D_{F}=300\)
(y-axis) as a function of \(<F>\)(x-axis),
circles denote simulations ending in an \(\alpha\) situation and squares denote
simulations ending in a \(\gamma\)
situation. The change in behavior is abrupt and occurs in the 0.55-0.61 region.
Simulations in that parameter region resulted in \(\beta\) situations.
}
\label{fig:phase300}
\end{figure}

In fig(\ref{fig:phase300}) we can see an abrupt phase-transition between
\(\alpha\) states to \(\gamma\) states for high \(D_{F}\) values around \(<F>=0.61\).
We shall give a theoretical explanation to this phase change.
For high \(D_{F}\) values \(\mathbf{F}\) reactants take shorter times to pass
between different parts of the simulation space. Therefore we can assume that \(\mathbf{A}\) reactants are
affected by all \(\mathbf{F}\) reactants in the simulation space and the
mean-field approximation:
\begin{equation}
\label{eq:meanapprox}
F=<F>
\end{equation}

is valid.

Assuming (\ref{eq:meanapprox}) we are now looking for solutions to:
\begin{equation}
\label{eq:constantF}
\frac{\partial{}\rho_{A}}{\partial{}t}=D_{A}\nabla^{2}\rho_{A}+k_{1}\rho_{A}\cdot{}<F>-k_{2}\rho_{A}^{2}-k_{3}\rho_{A}
\end{equation}

Keeping in mind that \(\mathbf{A}\) can take only integral values and dropping
the diffusion term from (\ref{eq:constantF}) we find that:
\begin{eqnarray}
\label{eq:solution_node}
<F>=\frac{k_{2}+k_{3}}{k_{1}}
\nonumber\\
\mathbf{A}=1
\end{eqnarray}

is an unstable node fixed point of (\ref{eq:constantF}). This leads to predict
the death of \(\mathbf{A}\)s for \(<F>\) smaller than  \(\frac{k_{2}+k_{3}}{k_{1}} \) and that
\(\mathbf{A}\)s fill up the space for \(<F>\) larger than  \(\frac{k_{2}+k_{3}}{k_{1}} \).

In the case of the simulated set of reactions (\ref{eq:move_F}) we have:
\begin{equation}
\frac{k_{2}+k_{3}}{k_{1}}\approx0.61
\end{equation}

The preceding argument explains the sharp transition between \(\alpha\) and
\(\gamma\) states for the high \(D_{F}=300\) value as can be seen in fig(\ref{fig:phase300}).

One might be tempted to think that this mean-field argument will be sufficient
in explaining the emergence of grazing at low \(D_{F}\) values: for low
diffusion coefficients persistent spatial discrepancies in \(\mathbf{F}\)'s concentration
prevail throughout the simulation space, some areas would be affected from a
local \(\mathbf{F}\) concentration larger than 0.61 and would sustain a
population of \(\mathbf{A}\)s and some areas with a lower local \(\mathbf{F}\)
concentration would be empty of \(\mathbf{A}\)s. These spatial discrepancies
change with time causing our clusters to move across the simulation space.  
If this was the sole mechanism leading to the grazing behavior one would expect
that the \(\beta\) simulation results would appear for low \(D_{F}\) values
equally distributed on both sides of the \(<F>\)=0.61 asymptote. Clearly,
fig(\ref{fig:phase}) shows us that this is not the case, the \(\beta\)
situations are concentrated around lower and lower \(<F>\) values as \(D_{F}\) decreases.
Therefore we should search for another mechanism in order to explain the
behavior seen in fig(\ref{fig:phase}). For lower \(D_{F}\) values, \(D_{A}\) is not
negligible and clusters are not only supported by an influx of \(\mathbf{F}\)
reactants, but are also supported by their own ability to move and ``find''
areas of high \(\mathbf{F}\) concentration in their surroundings. This
mechanism of ``searching'' for \(\mathbf{F}\) reactants should be part of an
explanation for the behavior seen in fig(\ref{fig:phase}) at low \(D_{F}\) values.

\section{Microscopic simulation of the Gray-Scott model}

\subsection{The Gray-Scott model - a pattern formation mechanism}
The Gray-Scott model\cite{gray_8} was first designed as a model of glycolysis
and as in the simplest form of Turing\cite{turing_7} pattern formation it involves two
reactants one of them enhancing the auto-catalysis of the other, but the
geometrical patterns resulting from this model are different from the ones
observed in the Turing pattern case and unlike the Turing pattern case pattern formation occurs when diffusion
coefficients are equal as well. The model is given by:
\begin{eqnarray}
\label{eq:gray-scott}
\frac{\partial\rho_{C}}{\partial{}t}=
\nabla^{2}\rho_{C}-\rho_{C}\rho_{A}^{2}+F(1-\rho_{C})\quad ,
\nonumber\\
\frac{\partial\rho_{A}}{\partial{}t}=
D_{A}\nabla^{2}\rho_{A}+\rho_{C}\rho_{A}^{2}-(F+k)\rho_{A}
\end{eqnarray}

 where \(\rho_{A}(\vec{x},t)\) and \(\rho_{C}(\vec{x},t)\) are the concentration
fields for two chemical reactants \(\mathbf{A}\) and \(\mathbf{C}\)
respectively. The terms \(F(1-\rho_{C})\) and \(-(F+k)\rho_{A}\) in (\ref{eq:gray-scott}) describe the
system as being in contact with an external reservoir
kept at \(\rho_{C}=1\) and \(\rho_{A}=0\). Indeed (\ref{eq:gray-scott}) accepts
the solution:
\begin{eqnarray}
\label{eq:solution}
 \rho_{C}=1
\nonumber\\
 \rho_{A}=0
\end{eqnarray}

Consider the equations that result from (\ref{eq:gray-scott}) by dropping 
the diffusion terms. The above mentioned stable fixed point exists throughout
parameter space but for some range of \(\mathbf{F}\) and \(\mathbf{k}\)
 Pearson\cite{pearson_10} has found another fixed point. Fixing \(\mathbf{k}\), increase
or decrease of \(\mathbf{F}\) causes the second steady state to be lost. The
process of losing or gaining fixed-points as a function of the system's
parameters is called bifurcation. Looking at the non-diffusive system's
behavior under change in the external parameters can be very useful in order
to understand the general behavior of the diffusive-system under microscopic simulation.

Two types of bifurcations are usually distinguished - saddle-node bifurcation
and Hopf bifurcation.

In a saddle-node bifurcation either a fixed point appears and splits into two
fixed points or two fixed points become one and then disappear. The main feature
of such bifurcations is the nature of the single fixed point at the
bifurcation. The linearized system must have one zero eigenvalue and one
non-zero eigenvalue at that point. In other words if
\begin{equation}
\label{eq:linearized_system}
\frac{\partial\vec{f(t)}}{\partial{}t}=\mathbf{M}\vec{f(t)}
\end{equation}

is the linear expansion of the  system around the fixed point, with
 \(\mathbf{M}\) being the 2x2
 coefficient matrix, then
at the bifurcation point \(det(\mathbf{M})=0\) and \(tr(\mathbf{M})\neq0\).

Hopf bifurcation occurs when an unstable (stable) focus goes through the
creation of a limit cycle and becomes stable (unstable). At the bifurcation
point both eigenvalues  are purely imaginary and the Real part of the
eigenvalues is positive (negative) before the bifurcation point and negative
(positive) after the bifurcation point.

In the case of our system, given \(\mathbf{k}\) the second fixed point is lost
through saddle-node bifurcation as \(\mathbf{F}\) is increased and by Hopf bifurcation as
\(\mathbf{F}\) is decreased.

Pearson \cite{pearson_9} considers
\(\rho_{C}\) as the density of a liquid fuel and \(\rho_{A}\) as a temperature field. Fuel is
constantly fed from an external reservoir kept at a constant concentration.
The fixed point in (\ref{eq:solution}) is stable therefore small perturbations in
the temperature field around the zero will tend to die out and return back to the
zero, but what if the perturbation was big enough (a match is thrown into the
fuel) to cause the term \(\rho_{C}\rho_{A}^{2}\) to be significant?

Here the auto-catalytic nature of (\ref{eq:gray-scott}) would cause temperature and
fuel consumption to increase inside the ignited area and the fire starts
spreading across space, leaving regions with low concentrations of fuel. Given
the feed parameters \(\mathbf{F}\) and \(\mathbf{k}\) the behavior is
controlled by the value of \(D_{A}\). Large values of \(D_{A}\) would cause fast moving fire wavefronts and
small values of \(D_{A}\) cause stable standing spots of fire fueled by the
fast moving \(\mathbf{C}\). One of the more interesting patterns arising from
(\ref{eq:gray-scott}) is that of replicating spots predicted in simulations
\cite{pearson_10} and confirmed in experiment \cite{experimental_spots}. Given the right parameter 
values an initial large perturbation breaks up into standing spots, the fuel
concentration at the middle of the spot decreases and causes the spot to divide
into two smaller expanding replicas and so on.

 Analytic spot-like solutions to (\ref{eq:gray-scott}) were found in
 \cite{pearson_9} but only in the one-dimensional case and in the
 \(D_{A}<<1\) limit.

The different two-dimensional patterns resulting from the Euler integration of (\ref{eq:gray-scott})
were classified in \cite{pearson_10}.

\subsection{Reproducing the replicating spots phenomenon in microscopic simulations.}
The following equations :
\begin{eqnarray}
\label{eq:PDEspots}
\frac{\partial\rho_{C}}{\partial{}t}=
2\cdot10^{-5}\nabla^{2}\rho_{C}-\rho_{C}\rho_{A}^{2}+0.018(1-\rho_{C})
\nonumber\\
\frac{\partial\rho_{A}}{\partial{}t}=
10^{-5}\nabla^{2}\rho_{A}+\rho_{C}\rho_{A}^{2}-0.074\rho_{A}
\end{eqnarray}

fall in the parameter space domain described by Pearson\cite{pearson_10} to produce the replicating
spots behavior. Pearson has used lattice-spacing of 0.01 and has started his
simulation in the trivial state: \(\rho_{A}=0\), \(\rho_{C}=0\) apart from the
perturbated square put at: \(\rho_{A}=0.25\), \(\rho_{C}=0.5\).
The reaction dynamics corresponding to equation (\ref{eq:PDEspots}) are:
\begin{enumerate}
\item \(\mathbf{A} + \mathbf{A} + \mathbf{C} \rightarrow  \mathbf{A}  + \mathbf{A} + \mathbf{A} (\mathbf{1})\)
\item \(\mathbf{A} \rightarrow \emptyset (\mathbf{0.74})\)
\item \(\mathbf{C} \rightarrow \emptyset (\mathbf{0.018})\)
\item \( \emptyset \rightarrow \mathbf{C} (\mathbf{0.018})\)
\end{enumerate}

The simulation space (of size 64x64 sites) is seeded with 1000 \(\mathbf{C}\)s and no \(\mathbf{A}\)s, except for the
perturbated square (of size 20x20 sites) that is seeded with 250 \(\mathbf{A}\)s per
cell and 500  \(\mathbf{C}\)s per cell. \(n_{0}\) was chosen to be \(10^7\) and
lattice spacing was chosen to be \(0.01\) (so a concentration of \(1\)
corresponds to \(1000\) reactants per lattice site).

\begin{figure}
\frame{
\vspace {0.01in}
\epsfxsize=3.1in
\epsfbox{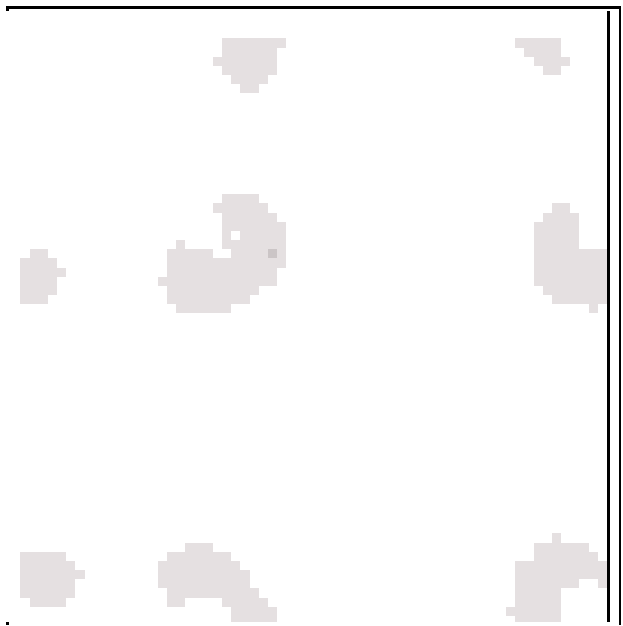}
}
\caption{
Replicating spots of \(\mathbf{A}\) created by the microscopic simulation.
 Spots in the process of dividing are shown.
}
\label{fig:spotsA}
\end{figure}

\begin{figure}
\frame{
\vspace {0.01in}
\epsfxsize=3.1in
\epsfbox{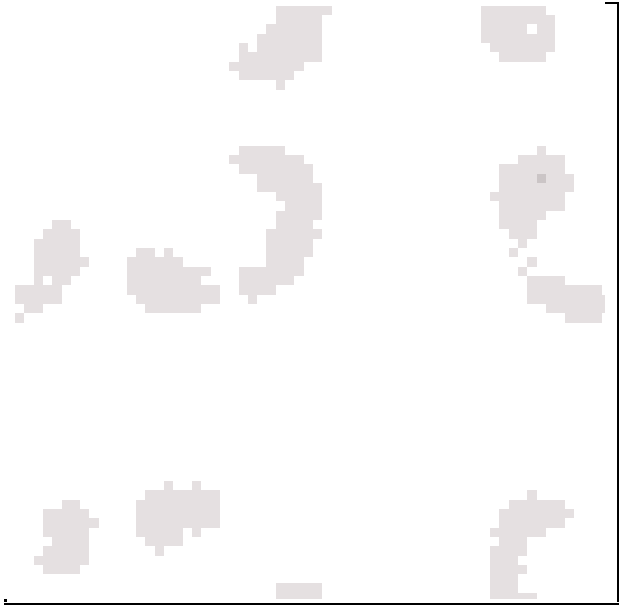}
}
\caption{
The same simulation as in 
fig(\ref{fig:spotsA}) at a later time, the original spots have split in two.
}
\label{fig:spotsC}
\end{figure}

\begin{figure}
\frame{
\vspace {0.01in}
\epsfxsize=3.1in
\epsfbox{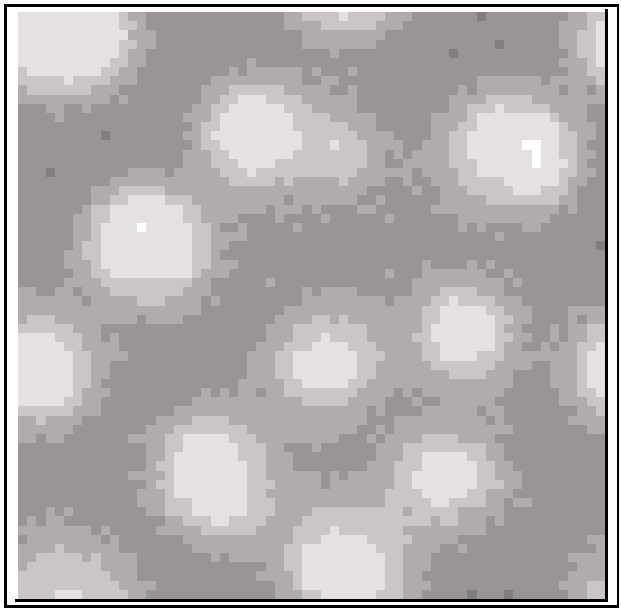}
}
\caption{
The same simulation as in 
fig(\ref{fig:spotsC}) and (\ref{fig:spotsA}) but at a later time. This figure
shows species \(\mathbf{C}\).
}
\label{fig:spotsD}
\end{figure}

\begin{figure}
\frame{
\vspace {0.01in}
\epsfxsize=3.1in
\epsfbox{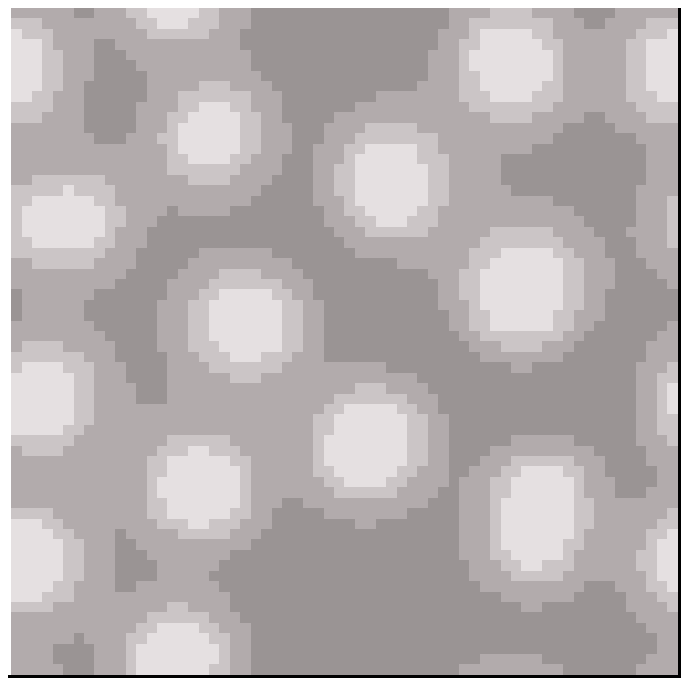}
}
\caption{
This figure shows results of PDE simulation of the replication spots. 
Again showing species \(\mathbf{C}\).
}
\label{fig:spotsE}
\end{figure}

Although we have been successful in reproducing the general replicating spots
behavior, a closer look at the results as shown in figures (\ref{fig:spotsC})
and (\ref{fig:spotsA}) shows that the randomness introduced in the microscopic
simulation has the effect of breaking the square symmetry preserved by the PDE
simulations. Pearson originally achieved this symmetry breaking by introducing noise
in the initial conditions, but the same effect is created by the microscopic simulation. 
Figure (\ref{fig:spotsD}) shows the results of the simulation at a later
time, replication has formed more spots. Figure (\ref{fig:spotsE}) shows the results of a PDE simulation of the
same system. We can see the similarity in the results of the two simulations,
 due to the fact that \(n_{0}\) is large. We shall see, in the next subsection,
that when other parameters are chosen (including \(n_{0}\)), the two systems
give crucially different results.

\subsection{The uniqueness of persistently dynamic reaction-fronts to
microscopic simulation. }
We now present an example in which microscopic simulations of the Gray-Scott model
 create different results than the ones obtained by the PDE approach. The
 phenomenon we are interested in is the presence persistent reaction fronts
 propagating through the simulation space. Much research has been devoted to
reaction fronts
\cite{oscillations}\cite{traveling_RNA}\cite{CIMA_1}\cite{CIMA_2}\cite{CIMA_3}.
Patterns which are
non-stationary in time and inhomogeneous in space at the same time are rare in a homogeneous
medium. We suggest a system where the reaction-fronts are spatially and
temporally non-stationary within a homogeneous medium (all the participating
reactants have non-zero equal diffusion rates, so the system is not equivalent
to a non-homogeneous medium system).
Our persistent reaction-fronts are
created by the following mechanism:  A localized small \(\mathbf{A}\) area 
consumes \(\mathbf{C}\) reactants in its surroundings and creates reaction-fronts of high
\(\mathbf{A}\) concentration propagating across the simulation space cleaning
areas from the presence of \(\mathbf{C}\)s. The reaction-front then runs out of
high \(\mathbf{C}\) concentration areas and decays, giving rise to the renewal
of \(\mathbf{C}\)s and new wavefronts produced by the surviving \(\mathbf{A}\)s and so on. We have
 simulated microscopically the following reaction-diffusion system: 
\begin{enumerate}
\item \(\mathbf{A} + \mathbf{A} + \mathbf{C} \rightarrow  \mathbf{A}  + \mathbf{A} + \mathbf{A} (\mathbf{8})\)
\item \(\mathbf{A} \rightarrow \emptyset (\mathbf{0.3})\)
\item \(\mathbf{C} \rightarrow \emptyset (\mathbf{0.02})\)
\item \( \emptyset \rightarrow \mathbf{C} (\mathbf{0.1})\)
\end{enumerate}

\begin{figure}
\frame{
\vspace {0.01in}
\epsfxsize=3.1in
\epsfbox{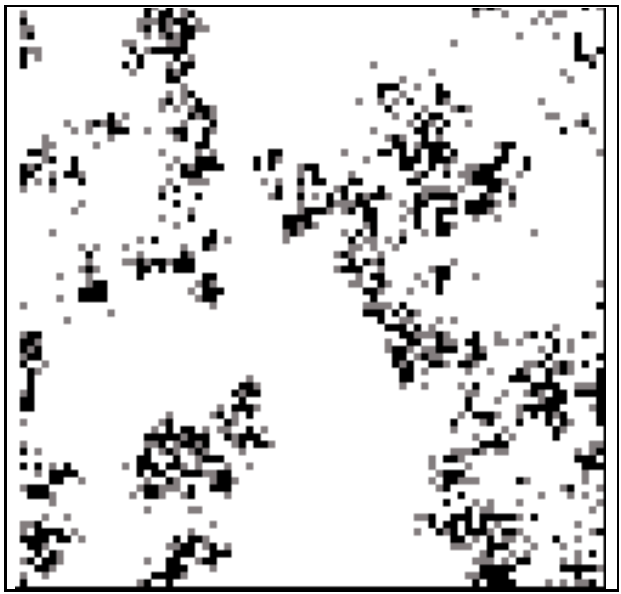}
}
\caption{
Clear high \(\mathbf{A}\) concentration reaction-fronts can be seen during microscopic simulations. These
reaction-fronts are persistent, being supported by surviving \(\mathbf{A}\) reactants.
}
\label{fig:mic_fronts}
\end{figure}

The diffusion coefficients are \(D_{A}=1\) and
\(D_{C}=1\). The simulation size chosen was 80x80. \(n_{0}=1\) and
\(\Delta{}x=1\). The simulation space was seeded with 1 \(\mathbf{C}\) per
site, and no \(\mathbf{A}\)'s were seeded except for a perturbed square of size
64x64 sites where 5 reactants of species \(\mathbf{A}\) per site were seeded.
The simulation shows persistent reaction-fronts of the type seen
in fig(\ref{fig:mic_fronts}).

\begin{figure}
\frame{
\vspace {0.01in}
\epsfxsize=3.1in
\epsfbox{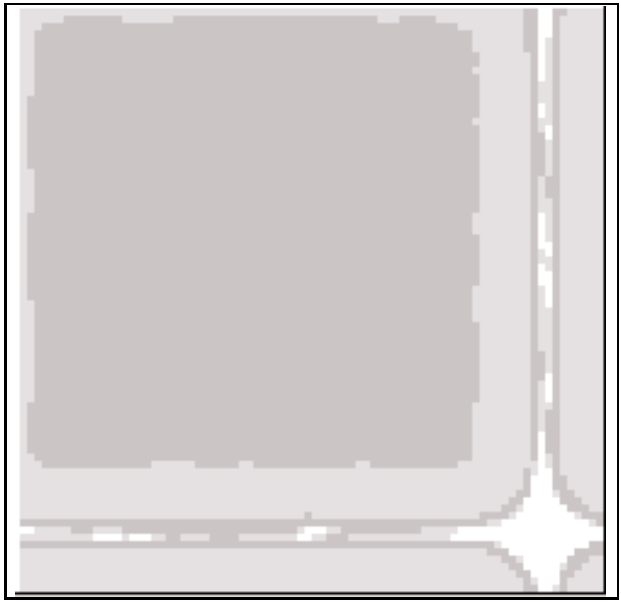}
}
\caption{
The initial reaction-front seen when integrating the corresponding PDE system
with \(\Delta{}x=1\) and \(n_{0}=1\). The reaction-front dies out and the
system is driven to the constant \(\rho_{A}=0\) steady-state.
}
\label{fig:pde_fronts}
\end{figure}
Using Euler Integration to simulate the system we get different results, the initial square perturbation does propagate in the form of
a reaction-front shown in fig(\ref{fig:pde_fronts}), but after the initial front cleans the space from high
\(\mathbf{C}\) concentration areas the front decays and gives way to the total
elimination of \(\mathbf{A}\)s from the simulation space. Unlike the
microscopic case in which the discretization and randomness leave behind some islands of
\(\mathbf{A}\) from which the next reaction-front can emerge, the Euler
Integration drives the system towards the \(\rho_{A}=0\)
steady-state. Introducing noise in the initial conditions does not change the
situation the system tends to smooth out these noises and is driven to the
constant steady state. 
 
We have tried to reproduce the persistent reaction-fronts pattern using Euler
Integration with no success, although it is plausible that the same behavior
will appear for high values of \(n_{0}\).

\section{Microscopic simulation of Marketing models}
\subsection{Using rate equations to describe the marketing of products}
The application of physical sciences methods to economic and financial
 research, nowadays common practice among the scientific community\cite{bouchaud_6}, was initiated by the
 work of Bachelier\cite{bachelier_3} and Mandelbrot\cite{mandelbrot_2}. The
 basic situation is similar to the other areas of research we have previously discussed, global ``macroscopic'' economic phenomena are generated by the underlying
 ``microscopic'' process of buy and sell. Under these circumstances it is
 natural to try and use microscopic simulation of market models in order to
 reproduce as an example we could take the
 spreading of steam engines or gunpowder starting from localized innovative centers
and sweeping across wide regions of the globe.  

When using microscopic simulation to describe product
marketing one can regard a lattice-space element in two different ways:
\begin{enumerate}
\item As a geographical region - neighborhood, town or country.
\item As a business entity - company or corporate.
\end{enumerate}
In the first case neighboring elements represent geographically close
regions, in the latter case they represent companies that are in business contact.    
The reactants could stand for any kind of valuable passing from one place or
business entity to another. One could have reactants representing
money, products, ideas, manpower or technology. Reactions could represent economic
processes in which valuables are transformed, lost or created. Diffusion stands
for the transfer of these valuables from one location or business entity to
another. Although the use of microscopic simulations in the context of marketing seems natural
enough there still are some flaws to point at and points to defend. The
microscopic simulation
is probabilistic in nature while some of the processes we try to
represent are deterministic. As we explained in previous sections rates
represent the occurrence rate of an underlying Poisson process for some of the
economic processes described we have no reason to think that they are
Poissonic in nature. Furthermore the microscopic simulation's lattice-space elements are always
connected with exactly 4 neighboring sites, thus disabling the simulation of
environments in which the degree of ``connectedness'' varies from site to
site. If we consider sites to be business entities as we have previously
suggested, we certainly would like to consider different amounts of
connectivity between business entities - a feature not supported by the
microscopic simulation system. 
 
\subsection {The ``Tamagotchi'' model}    
In collaboration with J. Goldenberg and D. Mazursky\footnote{Hebrew
University School of Business Management} we have devised the
following set of reactions:\footnote{Numbers at the right of reactions are the
corresponding reaction-rates.}
\begin{enumerate}
\item \(\mathbf{B} + \mathbf{C} \rightarrow  \mathbf{A}  + \mathbf{B} + \mathbf{B} (\mathbf{8})\)
\item \(\mathbf{A} \rightarrow \emptyset (\mathbf{0.7})\)
\item \(\mathbf{B} \rightarrow \emptyset (\mathbf{1.5})\)
\item \(\emptyset \rightarrow \mathbf{C}  (\mathbf{0.1})\)
\item \(\mathbf{C} \rightarrow \emptyset (\mathbf{0.02})\)
\end{enumerate}
The inspiration to these reaction dynamics comes from the wave of
``Tamagotchi'' games we have been subjected to during  a short period of 1996.
These reaction dynamics were simulated in a system of size 128x128 sites seeded
with 1 reactant of species \(\mathbf{C}\) only. A square of size 64x64 sites  was
seeded with 1 reactant of species  \(\mathbf{C}\)  per cell and 5 reactant of
species \(\mathbf{B}\) per cite. Other coefficients are: \(n_{0}=1\), \(\Delta{}x=1\).
\begin{itemize}
\item \(\mathbf{A}\) - represents a product (``Tamagotchi'').
\item \(\mathbf{B}\) - represents the idea or concept of the product.
\item \(\mathbf{C}\) - represents money.
\end{itemize}   
In this model neighboring locations should be interpreted in the geographical sense.
The only diffusing reactant is \(\mathbf{B}\) with diffusion coefficient
\(\mathbf{1}\) but since \(\mathbf{B}\)
represents an idea or concept the diffusion it performs is replicative. By
replicative we mean that instead of hopping from one cell to a neighboring
cell a copy of the original reactant is created and planted in the neighboring
cell. This replicative diffusion represents the fact that ideas or concepts
do not physically pass from one location to another, instead a location exposed
to a new idea or concept in a neighboring location creates its own copy of the
original idea. In terms of the microscopic simulation algorithm the only change is that when
executing the diffusion of \(\mathbf{B}\), a \(\mathbf{B}\) unit is added to the
target location but none is subtracted in the original location. The first
reaction represents the following process: A person exposed to the product
concept that is in possession of money spends the money on buying the product
and a new concept or idea of the product is ``born'' in that persons mind. The
second reaction represents the loss of products due to old age. The third
reaction represents the fact that ideas tend to be forgotten. The fourth
reaction represents an influx of money and the fifth reaction represents the
spending  of money on products other than \(\mathbf{A}\). The ``replicative''
diffusion that \(\mathbf{B}\) undergoes represents the influence ideas have on
neighboring locations.

\subsection{Wave fronts the ``Tamagotchi'' model}
During the simulations of the ``Tamagotchi'' model we have observed long
lasting wavefronts of high \(\mathbf{A}\) concentrations moving across the
simulation space. The process giving rise to such waves seems to be the
following: Starting with a small amount of \(\mathbf{A}\) in a \(\mathbf{C}\)
rich area the concepts created by this small \(\mathbf{A}\) concentration
propagate to neighboring sites and induce the decrease of \(\mathbf{C}\)
concentration and increase in \(\mathbf{A}\) concentration in these sites. The
areas to which this \(\mathbf{B}\) influence arrives then become \(\mathbf{C}\)
poor areas stopping the increase in \(\mathbf{A}\)'s concentration giving rise
to their decrease until \(\mathbf{C}\)'s concentration is high enough to
sustain the next wavefront.
\begin{figure}
\frame{
\vspace {0.01in}
\epsfxsize=3.1in
\epsfbox{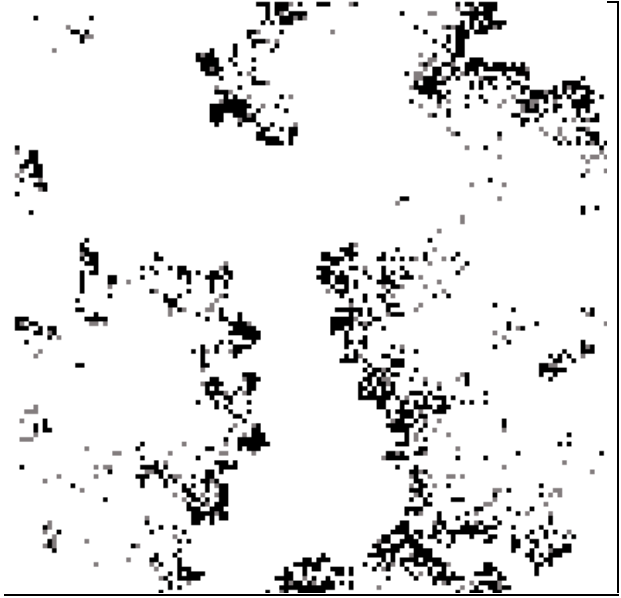}
}
\caption{
Snap shot of \(\mathbf{A}\)'s concentration clear wavefronts can be seen. Sites
with concentration above 2 are drawn black.
}
\label{fig:mesh131A}
\end{figure}

In order to estimate the propagation speed of the wavefronts we have used the
following correlation integral. Let \(\Omega\) denote our two-dimensional
reaction chamber and let \(\rho_{A}(\vec{x},t)\) denote the
concentration of \(\mathbf{A}\) at location \(\vec{x}\) and time
\(\mathbf{t_{0}}\) for \(\vec{x}\in\Omega\) and \(\mathbf{t_{0}}\geq\mathbf{0}\). For \(\mathbf{r}\geq\mathbf{0}\) let:
\begin{equation}
\label{eq:volume}
\textit{V}(\Omega)=\int_{\Omega}1d\vec{x}
\end{equation}

\begin{eqnarray}
\label{eq:correlations1D}
\mathbf{C}(r,t_{0},t)=\frac{1}{\textit{V}(\Omega)}\int^{2\pi}_0\int_{\Omega}\rho_{A}(\vec{x},t_{0})\rho_{A}(\vec{x}+
\left(
\begin{array}{c}
r\cos\theta\\r\sin\theta
\end{array}
\right)
,t_{0}+t)d\vec{x}d\theta - \nonumber\\ 
\frac{1}{\textit{V}(\Omega)^{2}}\int_{\Omega}\rho_{A}(\vec{x},t_{0})d\vec{x}\int_{\Omega}\rho_{A}(\vec{y},t_{0}+t)d\vec{y}
\end{eqnarray}

Given \(\mathbf{t}\geq\mathbf{0}\) and \(\mathbf{r}\geq\mathbf{0}\)
(\ref{eq:correlations1D}) is a measure for the
correlation between \(\mathbf{A}\)'s concentration at time \(\mathbf{t_{0}}\) and
\(\mathbf{A}\)'s concentration at time \(\mathbf{t_{0}+t}\) at locations with
distance \(\mathbf{r}\). If \(\rho_{A}\) is static \(C(r,t_{0},t)\)
is maximal for \(\mathbf{r}=\mathbf{0}\) for all \(\mathbf{t}\)s. If on the
other hand \(\rho_{A}\) is moving at speed \(\mathbf{v}\) we expect for a given
\(\mathbf{t}\) that C is maximal for \(\mathbf{r}\)=\(\mathbf{v}\mathbf{t}\).
\begin{figure}
\vspace {0.01in}
\epsfxsize=3.1in
\epsfbox{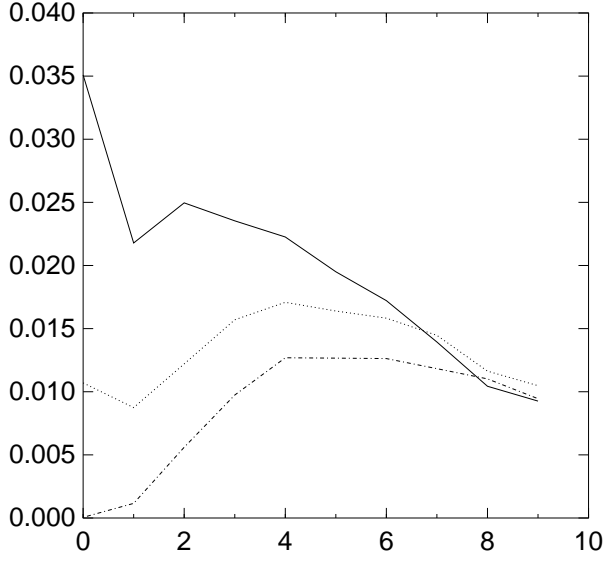}

\caption{
\(C(r,t_{0},t)\) plotted as a function of \textit{r} for \textit{t}=3,4,5, the
higher the plot the lower is \textit{t}. A clear 'hump' can be seen to advance in
\textit{r} as  \textit{t} increases, corresponding to the evolution of
the wavefront. 
}
\label{fig:speeds131A}
\end{figure}

\begin{figure}
\vspace {0.01in}
\epsfxsize=3.1in
\epsfbox{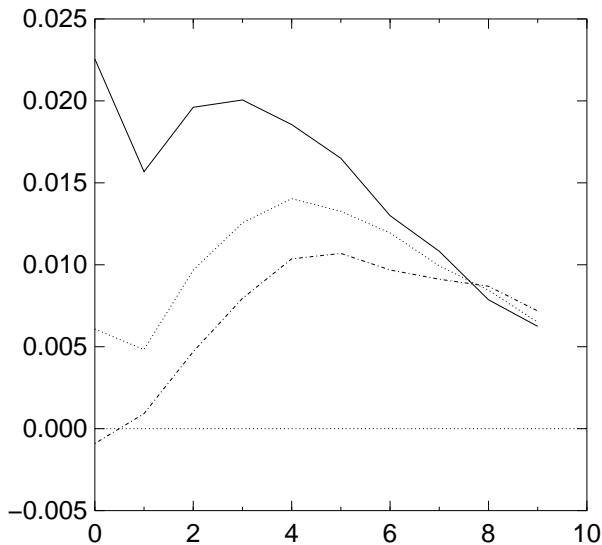}

\caption{
\(C(r,t_{0}+50,t)\) plotted as a f unction of \textit{r} for \textit{t}=3,4,5, the
higher the plot the lower is \textit{t}. The same 'hump' as in (\ref{fig:speeds131A})
although \(t_{0}\) has increased by 50.
}
\label{fig:speeds131B}
\end{figure}

In both (\ref{fig:speeds131A}) and (\ref{fig:speeds131B}) we can see an
increase of 2 in \(\mathbf{r}\) as \(\mathbf{t}\) increases by 2 giving rise
to an estimate of:
\begin{equation}
v\approx{}1
\end{equation}

When speeds are measured in units of lattice-sites per simulation-time.
The similar approximations due to the different choices of \(t_{0}\) indicate
the constant speed of advance of the wavefront.

\section{Acknowledgment}
We wish to thank Sorin Solomon for instructing us in this work. We thank Jacob
Goldenberg and David Mazursky their help in finding the application
to marketing. We also thank Nadav Shnerb for fruitful discussions.

\newpage
\bibliographystyle{plain}
\bibliography{cutdown}
\end{document}